# Optimal Cell Towers Distribution by using Spatial Mining and Geographic Information System

Prof. Dr. Alaa H. AL-Hamami
Amman Arab University for Graduate Studies
Amman, Jordan
Alaa_hamami@yahoo.com

Dr. Soukaena H. Hashem
University of technology
Iraq
soukaena_hassan@yahoo.com

*Abstract*— The appearance of wireless communication is dramatically changing our life. Mobile telecommunications emerged as a technological marvel allowing for access to personal and other services, devices, computation and communication, in any place and at any time through effortless plug and play. Setting up wireless mobile networks often requires: Frequency Assignment, Communication Protocol selection, Routing schemes selection, and cells towers distributions.

This research aims to optimize the cells towers distribution by using spatial mining with Geographic Information System (GIS) as a tool. The distribution optimization could be done by applying the Digital Elevation Model (DEM) on the image of the area which must be covered with two levels of hierarchy. The research will apply the spatial association rules technique on the second level to select the best square in the cell for placing the antenna. From that the proposal will try to minimize the number of installed towers, makes tower's location feasible, and provides full area coverage.

Keywords- Tower; Tower locations; Spatial mining; Digital Elevation Model; Database; and spatial association Rules.

## I. INTRODUCTION

Cellular telephony is the next and perhaps the most representative example of mobile communication systems. The cellular phone system is characterized as a system ensuring bidirectional wireless communication with mobile stations moving even at high speed in a large area covered by a system of base stations. The cellular system can cover whole country. Moreover, a family of systems of the same kind can cover the area of many countries. Initially, the main task of a cellular system was to ensure the connections with vehicles moving within a city and along highways. The power used by cellular mobile stations is higher than that used by the wireless telephony and reaches the values of single watts, for more details see references [1, 2].

Spatial Databases Spatial databases are databases that, in addition to usual data, store geographical information like maps, and global or regional positioning. Such spatial databases present new challenges to data mining algorithms. Spatial data mining is a process to discover interesting, potentially useful and high utility patterns embedded in spatial databases. Efficient tools for extracting information from spatial data sets can be of importance to organizations which own, generate and manage large spatial data sets, for more details see references [3, 4].

Association analysis is the discovery of what are commonly called association rules. The traditional problem is stated and solved in the following references [5, 6].

GIS, continuous surface such as terrain surface, meteorological observation (rain fall, temperature, pressure etc.) population density and so on should be modeled. Grid at regular intervals: Bi-linear surface

Polynomial is also used. Contour lines: Interpolation based on proportional distance between adjacent contours is used. TIN is also used. Profile: Profiles are observed perpendicular to an alignment or a curve such as high ways. In case the alignment is a straight line, grid points will be interpolated. In case the alignment is a curve, TIN will be generated [7, 8].

A DEM is a digital representation of topographic surface with the elevation or ground height above any geodetic datum. Followings are widely used DEM in GIS, see figure (1), [9, 10].

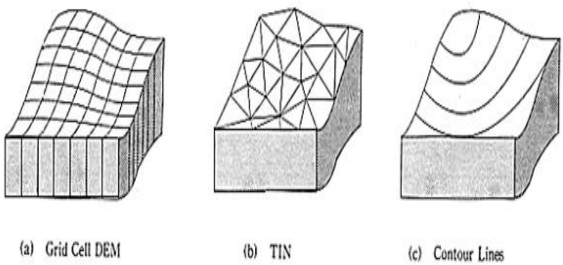

Figure 1: DEM models.

A DTM (Digital Terrain Model) is digital representation of terrain features including elevation, slope, aspect, drainage and other terrain attributes. Usually a DTM is derived from a DEM or elevation data. Several terrain features including the following DTMs. Slope and Aspect, Drainage network, Catchment area, Shading, Shadow and Slope stability, see figure (2), [11, 12].

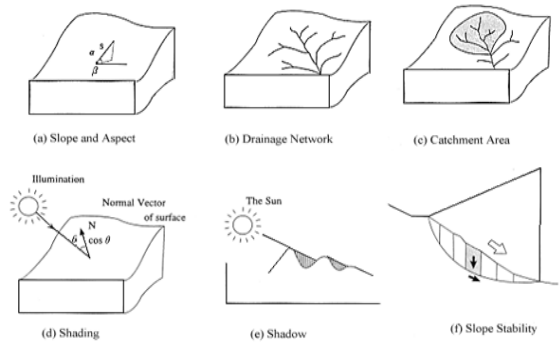

Figure 2: DTM models.

II. THE PROBLEM

Presume that you work for a cellular phone company that is interested in expanding the extent of its coverage on the country. You have been tasked with identifying the most suitable locations in the county for the placement of new cellular phone towers. As you might suspect there are many factors that govern the placement of cellular phone towers. Some factors are based on physical requirements, others on political and economical issues.

Now before we introduce the proposed system must clear some points related with tower distribution such as this problem is difficult to model so it is a NP-Complete. Also this problem has no exact Solution since it is multi-objective and constrained. So some sort of approximation is required to minimize number of antennas installed, to have practically feasible location, to provide full area coverage and finally to reduce the problem into a solvable solution.

III. THE PROPOSED SYSTEM

Distributing towers is difficult to model, so some approximation is required. This research introduces a suggestion to optimize the tower distribution and it will be explained in the following steps:

**First step (External Grid):**

Take image for the regions must be covered with mobile phones. Then applying GIS on the image using DEM for dividing the image to Grid, External Grid, each square in that grid will represent a cell, by a specific program depending on requirements of administrators of coverage the area with mobile phones, see figure (3). This was the first level with GIS using DEM on all the area must be covered.

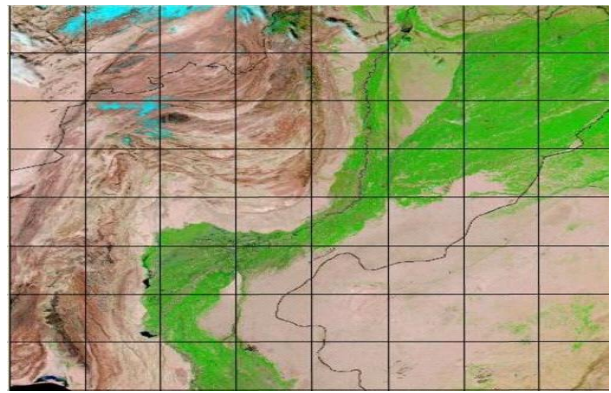

Figure 3: the over all area must be covered.

**Second step (Internal Grid):**

Each cell will represent a region must be covered by placing a tower so each cell in grid will also be divided into internal grid of square by the same specific program, see figure (4). To select the most suitable square for placing the antenna (The size of these squares depends on the coverage radius of the antenna used and here we suggest to use the omni-directional antenna which covers eight adjacent grid squares around it).

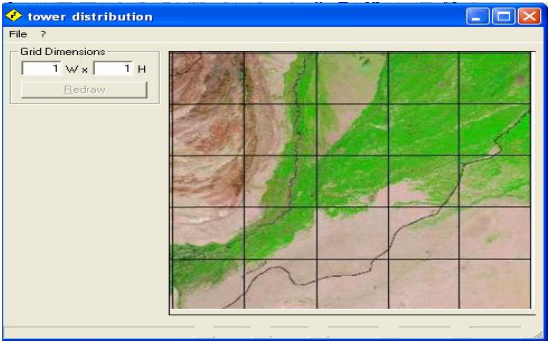

Figure 4: the cell which divided to squares.

**Third step (preprocessing the internal grid):**

Omni Directional Antennas are used to provide coverage in each cell. The tower must be placed in the most and most suitable location of the cell to make the covered area as good as coverage. To optimize the location of antenna we suggest the following:

- Data mining technique will be applied since we have the area (one of squares in the external grid) that will be divided into grid of squares of earth and the GIS for it. Both explicit and implicit relations and patterns among spatial objects in all squares will be extracted which are represent the presented area. The spatial and non-spatial attributes for all spatial objects in all the squares in the internal grid will be gotten.
- The following non spatial attributes (type, size, population, employment rate, etc. …..) and the spatial attributes are presented in the spatial database.

1. The first attribute will represent the spatial objects (O1, O2, O3, ……, On), and represents the identification of the transaction.
2. The second attribute will represent the type of the spatial object such as (town, road, ….) and will be represented as in the following codes:

**(Code of the type of the spatial object)**

1= town, 2 = road, 3 = river, 4 = sea, 5 = lake, 6 = mine, 7 = forest, 8 = bridge, 9 = highway, 10 = peak, 11= trough.

3. The third attribute will represent the size of the spatial object and will have the following codes:

**(Code of the size of the spatial object)**

1 = large, 2 = medium, 3=small.

4. The fourth attribute will represent the shape of the spatial object and will have the following codes:

**(Code of the shape of the spatial object)**

1 = point, 2 = line, 3=polygon.

5. The fifth attribute will represent the directions state: north of, south of, east of, west of, north west of, north east of, south west of, south east of. For more explanation see figure (1). These spatial attributes will have the following codes:

**(Code of the direction, Code of the related spatial object)**

(A, Oi) = (north of, Oi), (B, Oi) = (south of, Oi), (C, Oi) = (east of, Oi), (D, Oi) = (west of, Oi), (E, Oi) = (north east of, Oi), (F, Oi) = (north west of, Oi), (G, Oi) = (south east of, Oi), (H, Oi) = (south west of, Oi)

6. The sixth attribute will represent the Position state: disjoint, overlap, meet, covers, covered by. For more explanation see figure (5). These spatial attributes will have the following codes:

**(Code of the direction, Code of the related spatial object)**

(I, Oi) = (overlap, Oi), (II, Oi) = (meet, Oi), (IIIC, Oi) = (covers, Oi), (IV, Oi) = (covered by, Oi), (V, Oi) = (disjoint, Oi)

7. The seventh attribute will represent the distance between spatial objects, see figure (5).

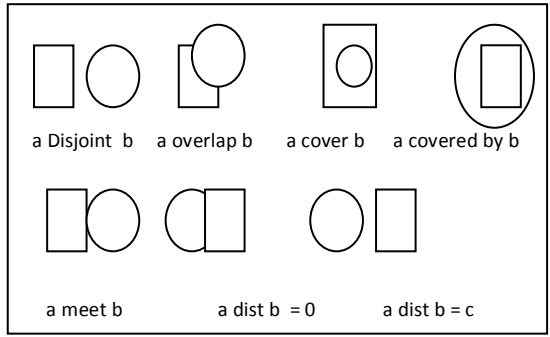

Figure 5: the position state spatial attribute..

8. For each square in the cell the proposed spatial database will be in the following design, Figure (6).

| Spatial object | Type | Size | Direction | Position | Dist | Population | Employment |
|---|---|---|---|---|---|---|---|
| O1 ……… ……… | 1 | 1 | (a, o2) | (I, o3) | O3 <50 km | High | high |

Figure 6: the design of the proposed spatial database.

**Fourth step (mining the internal grid):**

Now after building spatial database for each square in the cell we must mine this database to extract the patterns that will aid to select the best square to be the place of the antenna.

Since the proposed spatial database is usually relied upon alphanumerical and often transaction-based. The problem of discovering association rules is to find relationships between the existence of a spatial object (spatial or non spatial attributes) and the existence of other spatial objects (spatial or non spatial attributes) in a large repetitive collection. Association rules would give the probability that some objects attributes appear with others based on the processed transactions, for example large town ^ near to water → high population [90%], meaning that there is a probability of 0.9 that high population is found when the town is large and near water. Essentially, the problem consists of finding objects attributes that frequently appear together, known as frequent or large objects attributes-sets.

In this work (association rules) we find (spatially related) rules from the database of the square (3, 3). Association rules describe patterns, which are often in the database.

1. If type O1 = 4 (sea) and size O1 = 1 and direction O1 (south, o2 (type = 2)) then position O2 (type = 2) (intersect, O1) (s = 50%, c = 80%).

2. If type O1 = 4 (sea) then dist between O1 and (O2 (type =2) < 50 km (s = 50%, c = 90%).

**Analysis of extracted patterns:**

From the analysis we see:
- The resulted association rules from mining each square in cell are little and limited since the size of them is limited so usually the numbers of objects also limited.
- From analyzing the association rule of that square we see that this square presented nearly as a sea.

**Fifth step (classifying the squares of the internal grid):**

Now the system will classify the squares into two classes they are: first priority and second priority. The classification will be done according to the position of the square in the cell. The selection of position parameter for taken classifying because the basic threshold of used antenna was to be Omni Directional Antennas which has covers eight adjacent grid squares around it. So the rule of classification cell of n*n squares, each square has the position (x, y) will be as in the following:

*If square position was (1, any y) or (n, any y) or (any x, n) or (any x, 1)*

*then the square is **second priority class** else the square is **first priority class***

**Sixth step (selecting the optimal square):**

For selecting the optimal square follow the following steps;

1. From the results of spatial association rules on each square analyze these rules and from the analysis give the square will be assigned a ratio of goodness. The ratio of goodness is come from summation of:

- Class type (first priority present 100% and second priority present 50%). For

example the square which its position (3, 3) has 100% since it return to the first priority class.

- With suitability of the square (which come from the analyzing the association rules then the nature of the square will be understood. For example the rules extracted in step four with the square which its position (3, 3) which appear from the rules as a sea with little piece of road in this cases we think the suitability for placing antenna is nearly 50%.

**So the result for that square is 100% + 50% = 150%**

2. For each square repeat step one then compare which square has highest ratio of goodness will be chosen to be the best square for placing the antenna.

3. If the best ratio of goodness was not good enough then will begin in taken the squares in the class of second priority. Then apply point one and point two above in that sixth step. And take the best two corresponded squares to placing two antennas instead of one since that will guarantee the coverage of cell.

IV. CONCLUSIONS

From the suggestions and their results we get the following conclusions:

- Distributing and placing towers is a difficult problem to be modeled so the presented work was being as approximation for an optimal solution.

- Using GIS and DEM especially DTM making the division of area more accurate and presenting the surfaces of the square in more precious.

- Building spatial database as a flat database will make the spatial mining much more efficient that by reduce the mining to one level only so this will prevent the time and space consuming resulted in the previous work by extending the mining to multilevel.

- We proposed a novel approach for building a spatial database to accommodate all the necessary requirements for applying Association rules, and then extract all the patterns which help in distributing the towers.

- With the proposed spatial database the extraction of association rule is could done by the traditional apriori algorithm without confusing, that make the proposed approach easy to use and understand by the administrators. Also the analysis step followed by extraction rules is easy because it depends on generalization and normalization determined by the miner.

- In this research the classifier will be built without the need to measure the entropy of each attribute only it depends on the position of the squares.